\documentclass[12pt]{article}
\usepackage{epsfig}
\begin{document}

\begin {center}
{\bf Comment on Liu et al: `Final state interactions in the
decays $J/\Psi \to VPP$'}

\vskip 4mm {D.\ V.\ Bugg\footnote{email:
david.bugg@stfc.ac.uk} \\[2mm]
{\normalsize\it Queen Mary, University of
London, London E1\,4NS, UK} \\[3mm]}
\end {center}
\date{\today}

\begin{abstract}
\noindent
Statements concerning Extended Unitarity (EU) are clarified.

\vskip 2mm

\end{abstract}
\vskip 4mm

The objective of this Comment is to expose points of uncertainty
and disagreement concerning Extended Unitarity (EU).
Liu et al. \cite {Hanhart} make several assertions criticising an
earlier publication of mine \cite {EU}, where fits to four sets of
data are compared with the form of Extended Unitarity 
formulated by Au, Morgan and Pennington \cite {Au}.
The principles should apply to the overlap of
any two resonances having the same quantum numbers. Both my
publication and that of Liu et al. concern $\sigma$ and $f_0(980)$
where they overlap near 1 GeV.
Here $\sigma$ is a shorthand for the broad $\pi \pi$
component going through a phase shift of $90^\circ$ in the mass range
900--940 MeV.

For $\pi \pi$ elastic scattering, a Breit-Wigner resonance has an
elastic scattering amplitude
\begin {eqnarray}
\rho_{\pi\pi} T_{\pi \pi \to \pi \pi} &=& \frac {M\Gamma _{\pi \pi}(s)}
{M^2 - s - m(s) - iM\Gamma_{tot}(s)}
=\frac {N(s)}{D(s)} \\
m(s) &=&\frac {1}{\pi}P\int \frac {M\Gamma_{tot}(s')ds'}{s' - s};
\end {eqnarray}
$\rho$ is $\pi \pi$ phase space and N(s) is real. 
Watson's theorem \cite{Watson} amounts to the fact that the Breit-Wigner
denominator $D(s)$ and the associated pole are universal.
The numerator is not universal.

Fig. 1(a) shows the $\pi \pi$ mass projection from BES II data for
$J/\Psi \to \omega \pi ^+\pi ^-$ \cite {Ablikim}.
The peak near 500 MeV may be fitted with the $\sigma$ pole taking
$N(s) = 1$, unlike elastic scattering where $N(s)$ is constrained
by unitarity to be $M\Gamma_{\pi \pi}(s)$.
There may be a small dip in the data due to $f_0(980)$ in the bin just 
above the $KK$ threshold.
The $\sigma$ amplitude has a different $s$-dependence to elastic
scattering and the relative magnitude of $f_0(980)$ to $\sigma$
is also different.

\begin{figure}[htb]
\begin{center}
\vskip -12mm
\epsfig{file=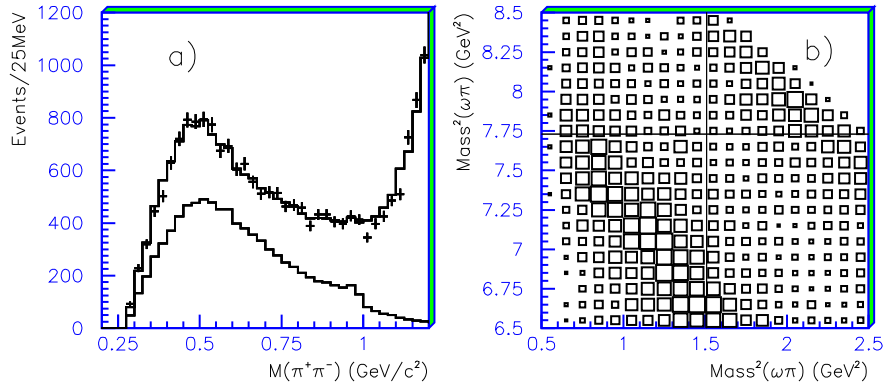,width=12cm}
\vskip -6mm
\caption
{(a) The $\pi \pi$ mass projection for BES II data on
$J/\Psi \to \omega \pi ^+\pi ^-$; the upper histogram
shows the fit to EU in the form of AMP, and the lower
histogram the $\pi \pi$ S-wave intensity;
(b) an enlarged view of the corner of the Dalitz plot;
cross-hairs show the point of intersection of $f_0(980)$
with the strong $b_1(1235)$ component.}
\end{center}
\end{figure}

Let us denote the $\pi \pi$ production amplitude by $A_k(s)$,
where $k$ refers to channels $\pi \pi$, $KK$, etc. 
Liu et al. use the relation
\begin {equation}
\rm {Im}  (A_i) = -\rho_k T^*_{ik}A_k,
\end {equation}
where 
$T_{ik}$ is the $2 \to 2$ amplitude.
The form of EU used by Liu et al. is that the phase of $T$ is identical
to that of elastic scattering.
They do not specify if this is the phase shift $\delta$ or the phase
$\phi$ measured from the bottom of the unitarity circle.
These differ above the inelastic threshold.
The Omn\` es relation \cite {Omnes} connects the magnitude of the
elastic scattering amplitude to $\phi$ using analyticity, so
$\phi$ appears to be the most relevant angle and will be used here.

The {\it critical}
assumption of EU is that the relative phases of $\sigma$ and $f_0$ in $T$
must be identical in production reactions and elastic scattering.
This is an assumption going beyond Watson's theorem \cite {Watson}.

In elastic scattering, $\sigma$ and $f_0$ amplitudes are constrained below
the $KK$ threshold to move around the unitary circle on the Argand diagram.
In Cern-Munich data \cite {Hyams}, it is obvious by eye that phase shifts
of $\sigma$ and  $f_0(980)$ add to a good approximation near 1 GeV.
This additivity of phases may be accomodated by multiplying
$S$-matrices $\eta \exp (2i\delta)$, though it is an open
question whether $\eta$ parameters should multiply.
For present considerations concerning the overlap of
$\sigma$ and $f_0$, this is not critical, since the
inelasticity of the $\sigma$ rises fairly slowly at the
$KK$ threshold because the $\sigma$ has a large $\pi \pi$ width.

For a production process the relative magnitudes of $\sigma$ and
$f_0$ are different, because of different matrix elements 
connecting initial and final states.
It is not obvious that $T$ of different resonances must combine in
the same way as for elastic scattering.
In a $2\to 2$ process, $\sigma$ and $f_0$ amplitudes combine
to make an overall amplitude $T$ which obeys the relation
\begin {equation}
{\rm Im} \, T = T T^*.
\end {equation}
In a $1	\to 3$ or $2 \to 3$ process, the boundary condition for
the initial state is different. 
Why can $f_0(980)$ not be completely absent in some production
processes? 
In that case, Eq (4) is still valid for the $\sigma$ amplitude
{\it alone}, but EU demands that the phase of $f_0(980)$ still appears 
in the production process despite its magnitude being absent.
Note that the process $J/\Psi \to \omega \pi ^+\pi ^-$ accounts
for only $1\%$ of all $J/\Psi$ decays.

These questions were much discussed in the early 1960s and led
to the isobar model, where each pole is multiplied by
complex coupling constants and amplitudes are added, not
multiplied.
For experimentalists, this is a simple form with which to parametrise
data, except possibly in the small region where they
overlap and EU could play a role.
Even there it parametrises average phase differences between
resonances.
In the production process, the matrix element involves an
unknown integral between the initial and final-state
wave functions.
This is what Liu et al. seek to model.
There is no dispute that multiple scattering processes do 
affect phases.

The Omn\` es relation, based on analyticity, allows the magnitude
of the $\pi \pi$ amplitude to be derived from phases $\phi$.
A subtlety is that the magnitude can be multiplied by a 
polynomial.
This polynomial accomodates, for example, a form factor arising
from the overlap integral between initial and final states.
Liu et al.  note this polynomial, but do not give details of
whether or how they use it.

On Fig. 1(a). the magnitude of the $\sigma$ amplitude is not the
same as for elastic scattering. My approach, consistent with the
data within errors, is to take $N(s) = 1$. 
Liu et al. may use instead an explicit polynomial
relating elastic scattering and production, but they do not say.
They derive relative magnitudes of $\sigma$ and $f_0$ from 
equations given in earlier work by
L\" ahde and Meissner \cite {Lahde}.
This takes $\sigma$ and $f_0$ to be $\bar nn$ and $\bar ss$
linear combinations, though results would be similar assuming
4-quark compositions (apart from SU3 Clebsch-Gordan
coefficients).
The matter of different SU3 Clebsch-Gordan coefficients is not
presently a critical issue for EU, though it would bear upon
the relation of $\sigma$ and $f_0$ to 2-quark, 4-quark and/or
meson-meson components in wave functions.
If the relative magnitudes of $f_0(980)$ and $\sigma$ are the same
in $J/\Psi \to \omega \pi\pi$ data and elastic scattering, the 
polynomial is 1 (except for standard form factors), but if the
relative magnitudes are different, a polynomial is required  peaking 
at the mass of $f_0(980)$ and reproducing its shape. 
What does this polynomial mean?

From the point of view of the isobar model, this is inconvenient.
The idea of the isobar model is that form factors should have small 
and slowly varying effects, usually negligible.
Liu et al do not show the Argand diagram for their amplitude.
It needs to be exposed what polynomial $P(s)$ is being used, so the
reader knows what price is being paid to preserve EU. 

Paradoxes can arise in limiting cases.
Suppose the $f_0$ amplitude is strictly zero in a production process.
One then sees the $\sigma$ amplitude alone.
According to Watson's theorem, the production
amplitude must then have the phase of the $\sigma$.
The hypothesis of EU requires a polynomial which exactly cancels
the magnitude {\it and} phase of the $f_0$ in elastic scattering; 
but the cancellation of the phase is not allowed, therefore a zero
$f_0$ amplitude appears to be forbidden by EU. My opinion is that
this is a critical defect in the hypothesis of EU. 

Let us return to details of $J/\Psi \to \omega \pi \pi$ data.
There were two fits reported in the BES publication
\cite {Ablikim}.
The second does contain a large $f_0(980)$ signal.
However, this gives a fit worse in $\chi^2$ by over 200
compared to a free fit to $f_0(980)$.
For fine details, the reader should consult Ref. [2], but
Fig. 1 presents essential points.
Fig. 1(b) shows an enlargement of the corner of the
Dalitz plot where the $f_0$ crosses the strong $b_1(1235)$
band in the data; the intersection is shown by the 
cross-hairs.
If there were a strong $f_0$ component with the same
phase as $\sigma$, one should see the interference
dip between them as a diagonal band running at $45^\circ$
from top left to bottom right. It should show a
distinctive interference as it crosses the $b_1(1235)$
band centred on the vertical cross-hair. 
This interference should have a full width at half-maximum of
0.07 GeV$^2$ in $m^2(\omega \pi)$.
There is no sign of this interference with
$b_1(1235)$, $\sigma$ or $f_2(1270)$.
This is why the fitted $f_0(980)$ components is very
small in my fit to the data, Ref. [2].

Liu et al. do not have access to the Dalitz plot. 
Their Fig. 5  shows structure just below 1 GeV which seems to
require an $f_0$ at least as large as the $\sigma$ amlitude in
this mass region. Firstly it is not obvious how to reconcile this
with $J/\Psi \to \phi \pi ^+\pi ^-$ data where the $\sigma$ amplitude
is much smaller than the strong $f_0(980)$ peak. 
Secondly the precipitious drop in the amplitude fitted to $\omega 
\pi ^+ \pi ^-$ data just below the $KK$ threshold should generate a
cliff-like diagonal structure on Fig. 1(b). 
The fit of Liu et al. is inconsistent with the isobar model fit
shown as the dashed curve on Fig. 1(a).  
The broad diagonal band of Fig. 1(b) arises from interferences of
$\sigma$ with $b_1(1235)$ and $f_2(1270)$.

\begin{figure}[htb]
\begin{center}
\vskip -12mm
\epsfig{file=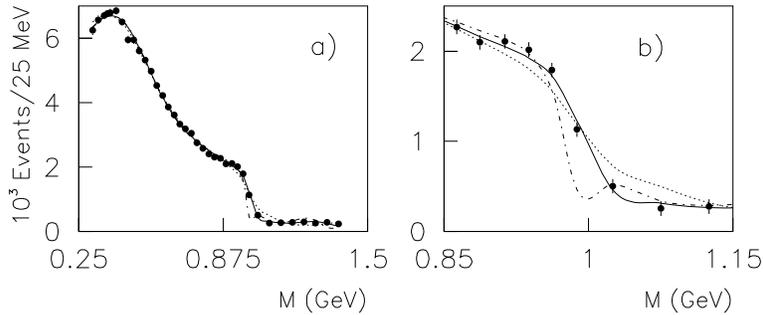,width=12cm}
\vskip -6mm
\caption
{(a) AFS data compared with the isobar model
fit; (b) enlargement near 1 GeV: the full curve is
the isobar model fit, the chain curve shows what appears in
elastic $\pi \pi$ scattering, and the dashed curve shows the isobar
model fit supplemented by a freely fitted $I=2$ $\pi \pi$ amplitude.}
\end{center}
\end{figure}

It would help the discussion if the BES II collaboration would make
publicly available the data set produced by myself and collaborators, 
the corresponding Monte Carlo data and the sidebin events used to
subtract the 14\% experimental background. The decay plane of the 
$\omega$ provides delicate information defining accurately the 
$b_1(1235)\pi$ signal, which then serves as an interferometer 
determining magnitudes and phases of amplitudes crossing it. 

A much clearer test of EU arises in data from the ISR \cite {ISR}
for central production of $\pi \pi$ in the process $pp \to pp(\pi \pi)$.
The central $\pi \pi$ pair
is far removed in rapidity from final states protons, and
there is no evidence for a Deck effect, i.e. resonant $p\pi$ or
$p\pi\pi$ combinations in the central region.
In this case, EU predicts that relative
phases of $\sigma$ and $f_0$ should be identical to $2 \to 2$
processes.
The data are easily fitted by the isobar model.
It requires relative intensities $f_0(980)/\sigma \sim 60\%$
of elastic scattering at the $KK$ threshold.
The fitted $f_0$ phase is $(57 \pm 7)^\circ$ below the
EU prediction.
Features of the fit are displayed in Figs. 9 and 10 of
Ref. \cite {EU}.
Fig. 2 summarises essential points.

Fig. 2(a) shows the isobar model fit (which includes a small
slowly varying form factor to reproduce the precise shape of
the sigma pole up to 0.8 GeV).
Fig. 2(b) shows details of the mass range near 1 GeV.
The full curve is the isobar model fit without any
$I=2$ $\pi \pi$ amplitude. The chain curve shows what is 
observed in elastic $\pi \pi$ scattering, suitably normalised.
These two curves are very different in the region of $f_0(980)$,
hence the phase difference required in the isobar model fit.
There could be an $I=2$ production amplitude. 
The dashed curve shows the result of fitting this freely. 
The fit is still poor and the $I=2$ amplitude is twice as 
large as in elastic scattering.
My view is that these data rule out EU in the form of Liu et al.

Morgan and Pennington \cite {Morgan} did manage to fit central
production data with EU but required an additional third-sheet
pole at $M=978 - i28 $ MeV.
Since then, data on $f_0(980)$ have improved,
and there is no  sign of this additional narrow third-sheet
pole in BES II data.

In Ref. [2], an empirical relation is presented which does
fit the ISR data. 
The assumption is made that the production
process can be considered as a black box. 
Then it is postulated that whatever linear combination of
$\sigma$ and $f_0$ is produced will rescatter according to the
usual relation $Im~T = TT^*$, i.e. the pions rescatter
as an isolated pair with magnitudes different from elastic
scattering.
If the asymptotic combination is written 
$T = T_\sigma + \beta T_{f_0}\exp ^{2i\Psi }$ within the elastic
regime, and $\beta$ is real, it can be shown that
\begin {equation}
2\Psi = 2\delta _\sigma - \delta _{f_0} + \sin ^{-1}(\beta \sin \delta_{f_0}).
\end {equation}
If $\beta \neq 1$, this is a different relation from the $2 \to 2$ process.
As $\beta$ decreases from 1, $\Psi $ falls rapidly.
In Ref. [2], the exact form $Im~T=TT^*$ is used, including effects of
coupling to $KK$.
Although Liu et al. dismiss this approach, it does have the merit of
correctly predicting the observed phase difference between $\sigma$ and
$f_0$ within errors.

No analyses so far include the dispersive cusp at the
$KK$ threshold in both $\sigma$ and $f_0(980)$.
The cusp originates from a discontinuity in slope of the
$KK$ amplitude at threshold.
This causes a corresponding discontinuity in slope of the
real part of the amplitude, hence the cusp \cite {Sync}.
I have made some preliminary fits to BES II data including
this cusp, and can offer some cautionary comments to
others who may try something similar.
In order to make the dispersion integral of Eq. (2) converge, a
form factor must be included in the $KK$ channel.
Whatever form factor is adopted, there is a narrow peak
in the real part of the $\pi \pi$ amplitude exactly at the
$KK$ threshold.
The narrow width implies  that mass resolution is of critical
importance in any comparison with data.
The width of the cusp is quite stable.
However, its height above zero real part does depend on
exactly what form factor is used.
Either this form factor must be predicted from a model of
$f_0(980)$  or the height
of the cusp must be treated as a variable in fitting data.
This source of uncertainty adds to difficulties in
deciding the nature of $f_0(980)$.

The advice to experimentalists is to use the isobar model,
which is simple and well defined.
It takes account empirically of phases such as Liu et al.
are trying to fit in a different way. 
If mass resolution is very good, it will be necessary to
include some $s$-dependence in $\Gamma _{\pi \pi}$
similar to the cusp.
Theorists can then compare the fit with models of $f_0$
and the cusp those models predict.
It may well be necessary to allow for $f_0 \to \eta \eta$.

In conclusion, the free fit made by Liu et al. to intermediate steps
in $J/\Psi \to \omega \pi ^+\pi ^-$ accounts for phases used in the
isobar model, but the magnitude of $f_0(980)$ signal they fit looks
inconsistent with Fig.l(b). The ISR data are in strong disagreement 
with EU.
In those data, there should be no significant perturbations
due to other resonances and the production mechanism is clean.


\begin{thebibliography}{99}

\bibitem{Hanhart}             
B. Liu et al., arXiv: 0901.1185

\bibitem {EU}                 
D.V. Bugg, Eur. Phys. J C54 (2008) 73 [arXiv: 0801.1908]

\bibitem {Au}                 
K.L. Au, D. Morgan and M.R. Pennington, Phys. Rev. D35 (1987)
1633

\bibitem {Watson}             
K.M. Watson, Phys. Rev. 88 (1952) 1173

\bibitem{Ablikim}             
M. Ablikim et al. (BES II Collaboration) Phys. Lett. B598
(2004) 149 [arXiv: hep-ex/0406038]

\bibitem {Omnes}              
R. Omn\` es, Nu. Cim. 8 (1958) 316

\bibitem {Hyams}              
B. Hyams et al., Nucl. Phys. B64 (1973) 134

\bibitem {Lahde}              
T.A. L\" ahde and U.-G. Meissner, Phys. Rev. D74 (2006) 034021
[arXiv: hep-ph/0606133]

\bibitem {ISR}                
T. $\AA $kesson et al., Nucl. Phys. B264 (1986) 154

\bibitem {Morgan}             
D. Morgan and M.R. Pennington, Phys. Rev. D48 (1993) 11851

\bibitem {Sync}               
D.V. Bugg, J. Phys. G: Nucl. Phys. 35 (2008) 075005 [arXiv: 0802.0934]
\end{thebibliography}
\end{document}